\pdfoutput=1
\documentclass[%
 reprint,
superscriptaddress,
 amsmath,amssymb,
 aps,
floatfix,
]{revtex4-2}

\usepackage{graphicx}% Include figure files
\usepackage{dcolumn}% Align table columns on decimal point
\usepackage[T1]{fontenc} % Use modern font encodings
\usepackage{siunitx}
\usepackage{verbatim}
\DeclareSIUnit{\belmilliwatt}{Bm}
\DeclareSIUnit{\e}{e}
\DeclareSIUnit{\dBm}{\deci\belmilliwatt}
\usepackage{amssymb}
\usepackage{xcolor}
\usepackage{braket}
\usepackage{bm}
\usepackage{booktabs}
\usepackage{hyperref}
\usepackage{graphicx} % Required for inserting images

\usepackage[utf8]{inputenc}
\usepackage[T1]{fontenc}
\title{Scientific Reports Title to see here}

\begin{document}
\preprint{APS/123-QED}

\title{All-electrical operation of a spin qubit coupled to a high-Q resonator}

\author{Rafael~S.\  Eggli}
\email{e-mail: rafael.eggli@unibas.ch, andreas.kuhlmann@unibas.ch}
\affiliation{Department of Physics, University of Basel, Klingelbergstrasse 82, CH-4056 Basel, Switzerland}

\author{Taras~Patlatiuk}
\affiliation{Department of Physics, University of Basel, Klingelbergstrasse 82, CH-4056 Basel, Switzerland}

\author{Eoin~G.\ Kelly}
\affiliation{IBM Research Europe-Zurich, Säumerstrasse 4,CH-8803 R\"uschlikon, Switzerland}

\author{Alexei~Orekhov}
\affiliation{IBM Research Europe-Zurich, Säumerstrasse 4,CH-8803 R\"uschlikon, Switzerland}

\author{Gian~Salis}
\affiliation{IBM Research Europe-Zurich, Säumerstrasse 4,CH-8803 R\"uschlikon, Switzerland}

\author{Richard~J.\ Warburton}
\affiliation{Department of Physics, University of Basel, Klingelbergstrasse 82, CH-4056 Basel, Switzerland}

\author{Dominik~M.\ Zumb\"uhl}
\affiliation{Department of Physics, University of Basel, Klingelbergstrasse 82, CH-4056 Basel, Switzerland}

\author{Andreas~V.\ Kuhlmann}
\email{e-mail: rafael.eggli@unibas.ch, andreas.kuhlmann@unibas.ch}
\affiliation{Department of Physics, University of Basel, Klingelbergstrasse 82, CH-4056 Basel, Switzerland}

\begin{abstract}

Building a practical quantum processor involves integrating millions of physical qubits along with the necessary components for individual qubit manipulation and readout. Arrays of gated silicon spins offer a promising route toward achieving this goal. Optimized radio frequency resonators with high internal quality factor are based on superconducting inductors and enable fast spin readout. All-electrical spin control and gate-dispersive readout remove the need for additional device components and simplify scaling. However, superconducting high-Q tank circuits are susceptible to crosstalk-induced ringup from electrical qubit control pulses, which causes fluctuations of the quantum dot potential and is suspected to degrade qubit performance. Here, we report on the coherent and all-electrical control of a hole spin qubit at $\SI{1.5}{\kelvin}$, integrated into a silicon fin field-effect transistor and connected to a niobium nitride nanowire inductor gate-sensor. Our experiments show that qubit control pulses with their broad range of higher harmonics ring up the tank when the control pulse spectrum overlaps with the tank resonance. This can cause a reduction of the readout visibility if the tank ringing amplitude exceeds the excited state splitting of the quantum dot, lifting Pauli spin blockade and thus leading to state preparation and measurement errors. We demonstrate how to circumvent these effects by engineering control pulses around the tank resonances. Importantly, we find that the ringup does not limit the spin coherence time, indicating that efficient high-Q resonators in gate-sensing are compatible with all-electrical spin control. 

\end{abstract}

\maketitle

\section{Introduction}
Scaling semiconductor spin qubit processors is a challenging endeavour but has recently accelerated \cite{Vandersypen2017, Hendrickx2021, Philips2022, Neyens2024}. The compatibility with industrial manufacturing \cite{Neyens2024} in concert with the small qubit footprint and operation at liquid helium temperatures \cite{Petit2020, Yang2020, Camenzind2022, Huang2024, Carballido2024} position silicon (Si) and germanium (Ge) spins at the forefront of these efforts. Recently, emphasis has been placed on the intricate interplay of device architecture and performance with qubit control schemes \cite{John2024, Carballido2024, Huang2024, Wang2024, Bartee2024},  heating \cite{Undseth2023a} and crosstalk effects \cite{Undseth2023, Cifuentes2023, Kelly2023}. A particular challenge lies in the integration of dedicated structures for spin readout and qubit control. Additional on-chip components beyond the qubit-defining gate electrodes such as single electron transistors for readout \cite{Hendrickx2021,  Philips2022, Borsoi2023, Huang2024} and microwave striplines \cite{Koppens2006, Huang2024} or micro magnets \cite{Philips2022, Neyens2024, Kawakami2014} for spin manipulation create undesirable overhead which impedes scalability, connectivity and qubit density.

Electrical spin control approaching $\SI{}{\giga\hertz}$ Rabi frequencies \cite{Froning2021, Wang2022, Fang2023} and $\SI{}{\micro\second}$ coherence times \cite{Hendrickx2021,Piot2022} has been achieved for hole \cite{Maurand2016, Camenzind2022} and electron \cite{Gilbert2023} Si spins and for holes in Ge \cite{Watzinger2018, Hendrickx2021, Froning2021}. The qubit is driven directly from a nearby gate thanks to the intrinsic spin-orbit interaction (SOI) \cite{Hendrickx2021,Camenzind2022,Froning2021,Gilbert2023}. Similarly, gate-dispersive sensing implements in-situ qubit readout \cite{Urdampilleta2019, West2019, Vigneau2023, Pakkiam2018} using gate electrodes as spin probes. The parasitic capacitance $C\mathrm{_{p}}$ of the sensing gate and other circuit components together with an off-chip inductor $L$ form a radio frequency (RF) resonator (tank circuit) whose resonance is sensitive to the qubit state \cite{Vigneau2023}. Combining all-electric qubit control and gate-sensing is a key step towards realizing a scalable architecture with high connectivity \cite{Crippa2019, Piot2022}. High readout sensitivities and speeds are achieved by increasing the internal quality factor $Q$ of the tank circuit which applies to all RF-based readout schemes \cite{Vigneau2023, GonzalezZalba2015}. An appealing approach utilizes superconducting nanowire \cite{Samkharadze2016, Niepce2019} or spiral \cite{Ahmed2018, Niegemann2022, Pakkiam2018} inductor with a high kinetic inductance to boost $Q$. 

However, to date, there are no reports of all-electrical coherent spin control in the presence of such a high-Q gate sensor. Recent experiments with niobium nitride (NbN) nanowire inductors have shed light on one potential reason for this gap \cite{Kelly2023}: capacitive crosstalk between qubit bond pads has been shown to promote resonator ringup if spectral components of the qubit drive pulses overlap with the tank circuit resonance. Hole spins with strong intrinsic SOI \cite{Hendrickx2021,Camenzind2022,Froning2021} and electrons with artificial SOI \cite{Kawakami2014,Philips2022,Neyens2024} may be particularly vulnerable to such resonator ringing. This is because the oscillating gate voltages couple to the spin \cite{Bosco2023,Carballido2024,Liles2021,Gilbert2023}.

Here, we demonstrate all-electrical coherent spin control at $\SI{1.5}{\kelvin}$ of a Si fin field-effect transistor (FinFET) hole spin qubit \cite{Geyer2021,Camenzind2022,Bosco2023,Geyer2024} with a NbN nanowire inductor connected to the qubit's plunger gate, forming a tank circuit with $Q\simeq1'000$ \cite{Kelly2023}. This configuration serves as a proxy of a qubit unit cell with minimal on-chip overhead. We first investigate the mechanism by which resonator ringing affects spin qubit operation. For this purpose, a low-Q tank circuit is attached to the qubit and coherently pumped at the tank resonance frequency. We then replace the tank with a high-Q resonator and observe crosstalk-induced ringup caused by qubit control pulses. The high-Q tank can be excited at its resonance frequency $f_0 = \SI{276.4}{\mega \hertz}$ and at higher modes in the $\SI{}{\giga\hertz}$ regime. Despite this ringup, we find conditions for which qubit control is successful without compromising coherence or transport-based readout contrast. 

We identify ways to navigate qubit control pulses around high-Q tank circuit resonances to prevent state preparation and measurement (SPAM) errors caused by resonator ringing. These strategies are universally applicable to other charge sensing approaches which feature high-Q inductors \cite{Ibberson2021,Oakes2023,Oakes2023a,Niegemann2022}. Low-frequency gate pulses (baseband pulses) generate resonator ringing over a broad range of pulse durations. Our results are thus particularly relevant for qubits which require complex sequences of baseband pulsing on many gates \cite{Struck2024,Weinstein2023,Loss1998, Wang2024}.

\section{Experimental setup and ringup hypothesis}

\begin{figure}[htb!]
    %\centering
    \includegraphics[width=0.47\textwidth]{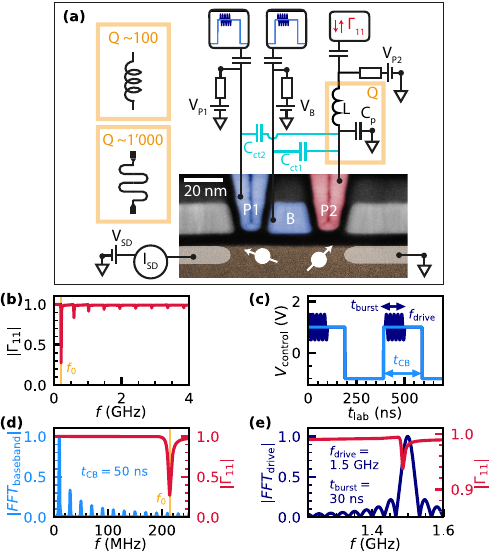}
    \caption{\textbf{Experimental setup and spectral overlap} \textbf{(a)} A false-colored transmission electron microscopy image of a FinFET 2-qubit device with circuit components is depicted. Two spins are accumulated in the silicon fin (brown) using DC voltages on two plunger gates (P1, P2) and a barrier gate (B). Drive lines are shown on P1 and B as well as a readout line connected to P2. Low-Q (high-Q) tank circuits are indicated by the symbol in the upper (lower) orange box on the left. Capacitive crosstalk (C$\mathrm{_{ct1/2}}$) leads to tank ringup from pulses applied to a drive gate. \textbf{(b)} The magnitude of the reflection coefficient $|\Gamma_{11}|$ of a high-Q tank circuit was simulated and features the tank resonance frequency $f\mathrm{_{0}}$ and higher modes. \textbf{(c)} A typical qubit drive sequence is composed of a symmetrical baseband square pulse (light blue) and a qubit drive pulse (dark blue). \textbf{(d)} The baseband pulse spectrum contains odd harmonics of order $n$ which are suppressed by $1/n$ and can match $f\mathrm{_{0}}$, causing ringup.  \textbf{(e)} The analytical spectrum of a single qubit drive burst exhibits a characteristic sinc-shape. Resonator ringup occurs via the higher tank harmonics because $f\mathrm{_{drive}}\gg f\mathrm{_{0}}$.}
    \label{fig:1Setup_Spectral_Overlap}
\end{figure}

The FinFET devices are fabricated using a self-alignment protocol yielding two gate layers with intrinsically perfect layer-to-layer alignment \cite{Geyer2021}. Fig.\ \ref{fig:1Setup_Spectral_Overlap} (a) shows a device with the key circuit components. The central barrier gate (B) controls the interdot tunnel coupling and has a larger lever arm $\alpha\mathrm{_{B}}\simeq0.2$ than the two plungers $\alpha\mathrm{_{P}}\simeq0.05$ (P1, P2). A generic experimental setup for simultaneous electrical spin control and gate-dispersive readout of one qubit comprises at least one drive and one readout line (see Appendix \ref{Supp:Circuit} for the detailed setup). On each line, a direct current (DC) voltage can be combined with a high frequency signal via a bias tee. In addition, the readout line features a tank circuit whose $Q$ is limited by internal losses. The orange boxes in Fig.\ \ref{fig:1Setup_Spectral_Overlap} (a) represent the two types of tank circuits used in the following: i) a low-Q ($Q\leq100$) tank featuring a commercial wire-wound surface mount inductor (upper  box), and ii) a high-Q ($Q\simeq1'000$) tank based on a NbN nanowire inductor (lower box) \cite{Kelly2023}. The bond pads ($100 \times \SI{100}{\micro\meter^2}$) of the FinFET device capacitively couple next neighboring gates with a crosstalk capacitance $C\mathrm{_{ct1}}\sim\SI{10}{fF}$ and second-next neigbouring gates with $C\mathrm{_{ct2}}\sim\SI{1.5}{fF}$ \cite{Kelly2023}. Qubit control pulses applied to a drive line thus leak to the neighboring gates such that the tank circuit rings if spectral components of the control pulse overlap with the tank resonance. 

We qualitatively investigate the conditions for which tank ringup is expected. The tank circuit has a fundamental resonance frequency $f\mathrm{_0}=\left(2\pi\sqrt{LC_{\mathrm{p}}}\right){}^{-1}$. Superconducting inductors can also have higher harmonics, as seen in the simulated tank reflection coefficient magnitude $|\Gamma_{\mathrm{11}}|$ (red) shown in  Figs.\ \ref{fig:1Setup_Spectral_Overlap} (b,d,e) , which can be explained by waveguide-like modes \cite{Davidovikj2017} (see Appendix \ref{suppfig:simulation} for a model). An idealized version of the most basic qubit control pulse is illustrated in Fig.\ \ref{fig:1Setup_Spectral_Overlap} (c) and consists of two components: i) the light blue baseband square wave of duration $t\mathrm{_{CB}}$ pulses the qubit from the readout/initialisation point to the manipulation point in gate space, where ii) the dark blue pulse at the drive frequency $f\mathrm{_{drive}}$ rotates the spin during the burst duration $t\mathrm{_{burst}}$. The qubit is read out by measuring the spin-dependent leakage current through the device using Pauli spin blockade (PSB) \cite{Camenzind2022}.

Both pulse components can cause tank ringing, but in different frequency regimes. Since the baseband fundamental frequency $f\mathrm{_{bb}} = \left(2t\mathrm{_{CB}}\right){}^{-1}$ is one to two decades lower than $f\mathrm{_0}$, higher harmonics of the baseband square pulse excite the tank if 

\begin{equation}
\label{eq:harmonics}
    f\mathrm{_0} =  n\cdot f\mathrm{_{bb}} = \frac{n}{2t\mathrm{_{CB}}}
\end{equation}

\noindent
is satisfied for odd integers $n$ \cite{Kelly2023}. An example spectrum is shown in Fig.\ \ref{fig:1Setup_Spectral_Overlap} (d) (light blue). Note that the magnitude of the higher harmonic components decreases with $1/n$. The spectrum of a qubit drive burst is a sinc pattern centered around $f\mathrm{_{drive}}$ as presented in Fig.\ \ref{fig:1Setup_Spectral_Overlap} (e) (dark blue). If the tank has higher modes, qubit driving can ring it up even for $f\mathrm{_{drive}}\gg f\mathrm{_{0}}$, which is typically satisfied for spin qubits (see Appendix \ref{Supp:Spectra} for details of the spectra).

\section{Experimental Results}
\subsection{Impact of tank ringing on a spin qubit}

\begin{figure}[hbt!]
    %\centering
    \includegraphics[width=0.47\textwidth]{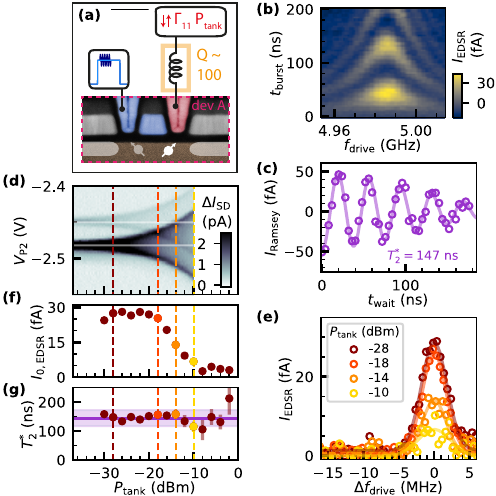}
    \caption{\textbf{Qubit coherence and readout with a pumped low-Q tank} \textbf{(a)} The setup with a low-Q tank circuit on P2 of device A is shown. \textbf{(b)} A Rabi chevron and \textbf{(c)} Ramsey trace of the qubit under P2 establish coherent spin control, yielding the spin dephasing time $T\mathrm{_2^*}=147\pm\SI{13}{\nano\second}$ and a Rabi frequency $f_\mathrm{Rabi}=\SI{12.5}{\mega\hertz}$. \textbf{(d)} Resonantly exciting the tank circuit at $f\mathrm{_{0}}$ and power $P\mathrm{_{tank}}$ through the input of the readout line causes broadening of the charge transitions. The qubit readout position is close to the ground state transition at $V\mathrm{_{P2}}\simeq\SI{-2.45}{\volt}$ and the excited state charge transition lies at $V\mathrm{_{P2}}\simeq\SI{-2.48}{\volt}$. At $P\mathrm{_{tank}}\simeq\SI{-18}{\dBm}$, the two charge transitions cross (red line). \textbf{(e)} Fitting the EDSR line in the coherence-limited regime $t\mathrm{_{burst}}\gg T\mathrm{_2^*}$ as a function of tank pump power yields \textbf{(f)} the readout contrast $I\mathrm{_{0,EDSR}}$ and \textbf{(g)} the qubit coherence $T\mathrm{_2^*}$, agreeing with the Ramsey experiment (purple line). A loss in readout contrast sets in only above the point where the charge transitions cross (red line) and reaches the noise floor at $P\mathrm{_{tank}}\geq\SI{-10}{\dBm}$ (yellow line). Despite this decrease in readout contrast, $T\mathrm{_2^*}$ remains constant within the error range. 
    }
    \label{fig:2LowQ_Tank_Coherence}
\end{figure}

We first investigate the effects of tank ringing on a hole spin qubit by directly pumping a low-Q tank circuit through the readout line as shown in Fig.\ \ref{fig:2LowQ_Tank_Coherence}. We expect no significant crosstalk-induced resonator ringup for the tank with $Q\leq 100$ due to the large $C\mathrm{_{p}}\simeq $\SI{1}{\pico\farad} \cite{Kelly2023} and because the decay of the excitation occurs on the time scale $t\mathrm{_{ringdown}}\simeq Q/f\mathrm{_0}\leq\SI{300}{\nano\second}$. The tank is connected to gate P2 of FinFET device A and we focus the following investigations on the spin which is closest to P2. This maximizes the impact of the tank ringing on the qubit. As shown in Figs. \ref{fig:2LowQ_Tank_Coherence} (b,c) the qubit can be coherently driven from P1 at $\SI{1.5}{\kelvin}$ with the spin dephasing time $T\mathrm{_2^*}=147\pm\SI{13}{\nano\second}$ at Rabi frequency $f_{\mathrm{Rabi}}=\SI{12.5}{\mega\hertz}$, in line with previous reports \cite{Camenzind2022}. 

Pumping the tank at power $P\mathrm{_{tank}}$ leads to a broadening of the quantum dot charge transitions as seen in Fig.\ \ref{fig:2LowQ_Tank_Coherence} (d) with several $\SI{}{\milli\volt}$ tank ringing amplitude \cite{Kelly2023}. We extract qubit properties by driving the qubit for a long time $t\mathrm{_{burst}}\gg T\mathrm{_2^*}$ and at low drive amplitude. In this regime, the EDSR linewidth is coherence-limited and we can extract $T\mathrm{_2^*}$ as well as the resonance amplitude $I\mathrm{_{0,EDSR}}$ (i.e. readout contrast) \cite{Kawakami2014,Hanson2007}. The fits and results are presented in Figs.\ \ref{fig:2LowQ_Tank_Coherence} (e, f, g). 

$I\mathrm{_{0,EDSR}}$ remains constant up to $P\mathrm{_{tank}}\simeq\SI{-18}{\dBm}$ (red line). This is exactly the power at which the ground (faint feature at $V\mathrm{_{P2}}\simeq\SI{-2.45}{\volt}$) and excited state (dark feature at $V\mathrm{_{P2}}\simeq\SI{-2.48}{\volt}$) cross in Fig.\ \ref{fig:2LowQ_Tank_Coherence} (d) due to broadening, indicating that the voltage oscillations on P2 are sufficient to overcome the excited state splitting. For even higher powers, $I\mathrm{_{0,EDSR}}$ decreases monotonically until reaching the noise floor above $P\mathrm{_{tank}}\simeq\SI{-10}{\dBm}$ (yellow line). Note that the $T\mathrm{_2^*}$ extracted from the linewidth remains constant within errors throughout this range, agreeing well with the Ramsey experiment (purple line) (see Appendix \ref{Supp:Analysis} for the fit functions). 

We conclude that resonator ringing is not limiting the coherence of our qubit. Intuitively, the crossing of the ground and excited state charge transition represents the point where state leakage out of PSB is possible during the readout and initialisation phase. The blocked spin can access the blockade-lifting excited state, thus incurring a SPAM error and lowering $I\mathrm{_{0,EDSR}}$.

\subsection{Crosstalk-induced ringup and qubit coherence}

\begin{figure*}[hbtp]
    %\centering
    \includegraphics[width=0.97\textwidth]{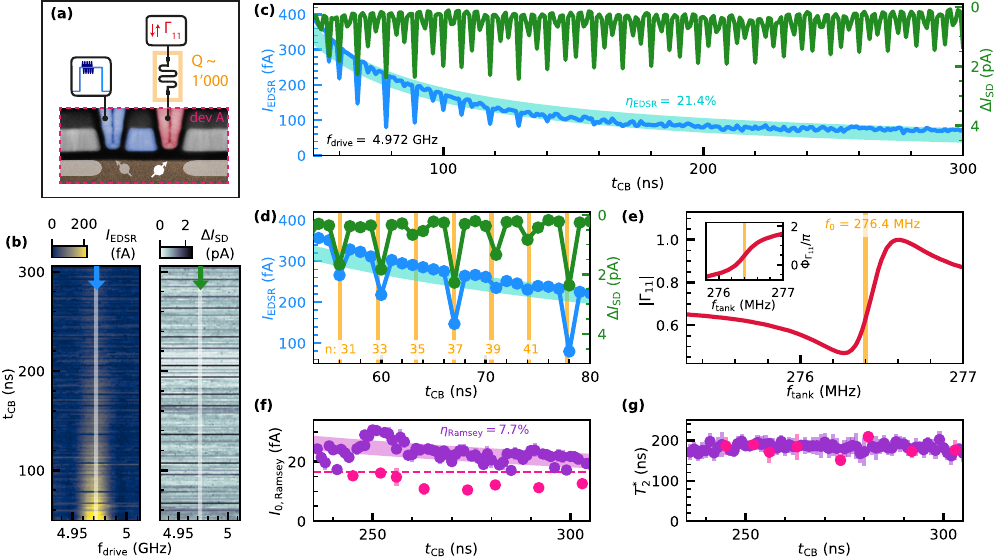}
    \caption{\textbf{High-Q tank ringup and qubit coherence} \textbf{(a)} Schematic setup with a high-Q tank circuit connected to gate P2 of device A. \textbf{(b)} The qubit resonance in $I\mathrm{_{EDSR}}$ (left panel) and offset current $\Delta I\mathrm{_{SD}}$ (right panel) was measured as a function of $t\mathrm{_{CB}}$ with $t\mathrm{_{burst}}$ chosen such that the spin is flipped on resonance. Line cuts of the two maps on the qubit resonance (arrows in \textbf{(b)}) are shown in \textbf{(c)} and, focussing on the first $\SI{30}{\nano\second}$, in \textbf{(d)}. $I\mathrm{_{EDSR}}$ (blue) shows a characteristic decay with $t\mathrm{_{CB}}$ which is captured well by the fit (turquoise), taking into account the experimental repetition rate $f\mathrm{_{bb}}$ and efficiency $\eta\mathrm{_{EDSR}}$. Additionally, $I\mathrm{_{EDSR}}$ dips significantly below this trend for specific $t\mathrm{_{CB}}$ values, accompanied by a sharp increase in $\Delta I\mathrm{_{SD}}$. These features are independent of $f\mathrm{_{drive}}$. From the tank resonance shown in \textbf{(e)} (phase in inset) the resonance frequency $f\mathrm{_{0}}$ is found. Odd harmonics of order $n$ of the baseband pulse match $f\mathrm{_{0}}$ for specific $t\mathrm{_{CB}}$ (orange lines in panel \textbf{(d)}) which agree excellently with the dips in $I\mathrm{_{EDSR}}$ and peaks in $\Delta I\mathrm{_{SD}}$ given the resolution limit of $\SI{1}{\nano\second}$. \textbf{(f)} The fitted amplitudes of Ramsey traces $I\mathrm{_{0, Ramsey}}$ as a function of $t\mathrm{_{CB}}$ follow a decay similar to $I\mathrm{_{EDSR}}$ with $\eta\mathrm{_{Ramsey}}$. Imposing a threshold (pink dashed line) identifies $t\mathrm{_{CB}}$ with significant ringing (pink). \textbf{(g)} Fitted $T\mathrm{_2^*}$ with colours corresponding to \textbf{(f)}.  
    }
    \label{fig:3Baseband_Ringup_T2}

\end{figure*}

We now swap the inductor on device A to a high-Q superconducting NbN nanowire with a nominal inductance of $L=\SI{1}{\micro\henry}$ \cite{Kelly2023}, connecting it to P2 as shown in Fig.\ \ref{fig:3Baseband_Ringup_T2} (a). The tank resonance depicted in Fig.\ \ref{fig:3Baseband_Ringup_T2} (e) with $f\mathrm{_0}=\SI{276.4}{\mega\hertz}$ implies a parasitic capacitance $C\mathrm{_p}=\SI{0.332}{\pico\farad}$. After the thermal cycle and re-bonding we operate the qubit in a very similar regime in gate voltage space but observe an overall reduced readout contrast (see Appendix \ref{Supp:Rabi_comparison} for a comparison of Rabi scans with the low-Q and high-Q resonators).

The left panel of Fig.\ \ref{fig:3Baseband_Ringup_T2} (b) shows the EDSR resonance of the qubit as a function of $t\mathrm{_{CB}}$, measured with $t\mathrm{_{burst}}=\SI{40}{\nano\second}$ which corresponds to a $\pi$ pulse on resonance. In the right panel, we depict the simultaneously measured source-drain offset current $\Delta I\mathrm{_{SD}}$ (for details on the measurement scheme, see Appendix \ref{Supp:Measurement_scheme}). Line cuts which were recorded with a higher integration time on the qubit resonance (arrows in (b)) are shown in Fig.\ \ref{fig:3Baseband_Ringup_T2} (c). Independent of tank ringup, $I\mathrm{_{EDSR}}$ is expected to decay with:

\begin{equation}
    \label{eq:I_EDSR}
    I\mathrm{_{EDSR}} \left(t_{CB}\right) = \eta\mathrm{_{EDSR}} f\mathrm{_{bb}} e = \eta\mathrm{_{EDSR}}\cdot\frac{e}{2t\mathrm{_{CB}}}
\end{equation}

\noindent
where the experimental repetition rate is $f\mathrm{_{bb}}=\left(2t\mathrm{_{CB}}\right){}^{-1}$, $e$ is the elementary charge, and the unit-less prefactor $\eta\mathrm{_{EDSR}}$ represents the efficiency of the EDSR experiment. Fitting Eq.\ (\ref{eq:I_EDSR}) to the $I\mathrm{_{EDSR}}$ trace yields $\eta\mathrm{_{EDSR}}= 21.4\pm0.4\%$ (turquoise curve in Fig.\ \ref{fig:3Baseband_Ringup_T2} (c) and (d)). The EDSR efficiency could approach $100\%$ but lies lower because of losses such as reservoir leakage.

The monotonous decay of $I\mathrm{_{EDSR}}$ is interrupted by sharp dips. These dips line up perfectly with peaks in $\Delta I\mathrm{_{SD}}$ as can be seen from Fig.\ \ref{fig:3Baseband_Ringup_T2} (d). Note the inverted current axis for $\Delta I\mathrm{_{SD}}$ to improve visibility of the correspondence. The $\SI{1}{\nano\second}$ resolution in $t\mathrm{_{CB}}$ is limited by the sampling rate of our control electronics. 

From the tank circuit resonance shown in Fig.\ \ref{fig:3Baseband_Ringup_T2} (e) and using equation \ref{eq:harmonics}, we can predict the $t\mathrm{_{CB}}$ for which the odd baseband pulse harmonics of order $n$ should match the tank resonance frequency. These are depicted as orange lines in Fig.\ \ref{fig:3Baseband_Ringup_T2} (c). The peaks in $\Delta I\mathrm{_{SD}}$ occur at almost every odd harmonic, but the dips in $I\mathrm{_{EDSR}}$ are more sparse.  

The amplitudes of the peaks in $\Delta I\mathrm{_{SD}}$ decrease for higher $t\mathrm{_{CB}}$, in line with the scaling of the baseband harmonics amplitude with $1/n$. $I\mathrm{_{EDSR}}$ is only expected to be reduced if the ringing amplitude exceeds the excited state splitting which explains the absence of a dip e.g.\ for $n=35$ because the harmonic lies in between two neighbouring datapoints whereas the dips are deep if a datapoint close-to perfectly matches a harmonic order (e.g.\ $n=\{31; 37; 42\}$). The dips in $I\mathrm{_{EDSR}}$ are less frequent and less deep for $t\mathrm{_{CB}}\geq\SI{100}{\nano\second}$ due to the reduced excitation power for higher $n$. Appendix \ref{Supp:Baseband_amplitude} presents the impact of the baseband pulse amplitude on the EDSR resonance. For $\Delta I\mathrm{_{SD}}$ on the other hand, the response is monotonous with ringing amplitude, explaining the abundance of peaks. The asymmetry of the tank resonance is an additional confounding factor when interpreting the dip amplitudes which is explored in more detail in Appendix \ref{supp:detuning_ringup}.

Finally, the effect of crosstalk-induced ringing on the qubit coherence is investigated. Ramsey experiments are performed for $t\mathrm{_{CB}}\geq\SI{235}{\nano\second}$, limited by the sum of $T\mathrm{_2^*}$ and the time required to perform two $\pi/2$ pulses ($t\mathrm{_{\pi/2}}=\SI{20}{\nano\second}$), yielding the Ramsey contrast $I\mathrm{_{0,Ramsey}}$ and $T\mathrm{_2^*}$. The results with standard errors of 20 repetitions are shown in Figs.\ \ref{fig:3Baseband_Ringup_T2} (f) and (g), respectively. 

The current amplitude decays according to a relation similar to Eq.\ (\ref{eq:I_EDSR}) (pink line fit in Fig.\ \ref{fig:3Baseband_Ringup_T2} (f)). We find the Ramsey efficiency to be $\eta\mathrm{_{Ramsey}}= 7.7\pm0.1\%$, significantly lower than $\eta\mathrm{_{EDSR}}$. This discrepancy may be due to the overall longer $t\mathrm{_{CB}}$ and the different pulse sequence with two $\pi/2$ pulses whose short burst duration causes more broadband excitation of the higher tank modes.

 Imposing a threshold of $I\mathrm{_{threshold}}=\SI{16.5}{\femto\ampere}$ (pink dashed line), we identify 8 points in the $I\mathrm{_{0,Ramsey}}$ plot which are clearly affected by ringup (pink). Comparing the fitted $T\mathrm{_2^*}$ for these $t\mathrm{_{CB}}$ values, no systematic deviation from the other datapoints (purple) is apparent. We thus conclude that the qubit coherence is not affected but SPAM errors reduce the readout contrast also for crosstalk-induced ringup. The difference in $T\mathrm{_2^*}$ as opposed to the the low-Q measurements may be due to the thermal cycle, causing rearrangements of charge fluctuators near the qubit and impacting the microscopic noise environment.

\subsection{Higher modes of the high-Q tank}

\begin{figure}[hbtp]
    %\centering
    \includegraphics[width=0.47\textwidth]{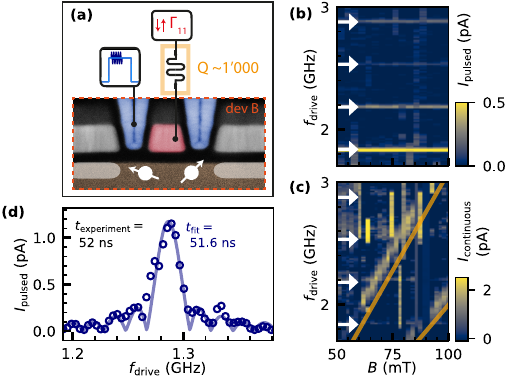}
    \caption{\textbf{Higher modes of the high-Q tank} \textbf{(a)} Schematic setup with the high-Q tank connected to the barrier gate of device B. \textbf{(b)} In the pulsed EDSR scan, horizontal lines (white arrows) which correspond to waveguide-like modes of the tank appear but no EDSR resonance.  \textbf{(c)} Identical scan as in \textbf{(b)} but implementing a continuous wave drive. Both qubit resonances are visible (orange lines are offset for better clarity). The continuous drive is spectrally very narrow, resulting in the faint indications of the higher tank modes (white arrows). \textbf{(d)} A high-resolution scan of one of the higher tank resonances exhibits the typical sinc-shape and its fitted $t\mathrm{_{fit}}=51.6\pm\SI{0.5}{\nano\second}$ agrees well with the experimentally chosen $t\mathrm{_{CB}}$.
    }
    \label{fig:4Haronics_Tank}
\end{figure}

In order to increase the sensor gate lever arm and thus achieve stronger coupling to the spin, a second FinFET, device B, was operated according to the schematic in Fig.\ \ref{fig:4Haronics_Tank} (a). The high-Q tank circuit was connected to the gate B, enabling charge sensing down to the last hole in earlier experiments \cite{Kelly2023}. This arrangement comes at the cost of higher susceptibility to crosstalk because the driving and the sensor gate bond pads are next-neighbors.   

Pulsed experiments where $f\mathrm{_{drive}}$ is swept against $B$ while applying the typical Rabi pulse scheme have repeatedly failed to record the EDSR resonance. A series of evenly spaced resonance lines appear as shown in Fig.\ \ref{fig:4Haronics_Tank} (b) (white arrows), which do not depend on the applied magnetic field. If instead the qubit is driven continuously the two EDSR resonances are observed (orange lines in Fig.\ \ref{fig:4Haronics_Tank} (c)) for otherwise identical conditions (see Appendix \ref{Supp:CW} for the measurement scheme). The spectrum of the continuous wave experiment approaches a delta peak at $f\mathrm{_{drive}}$, thus minimizing the resonator ringup to a narrow band around the higher modes (white arrows).

The detailed shape of one of the horizontal resonances in the pulsed scan is shown in Fig.\ \ref{fig:4Haronics_Tank} (d). Fitting the sinc pattern of the resonance, we find excellent agreement with the applied drive pulse duration. The relatively broad band drive pulse spectrally overlaps with a higher tank mode at $f_{\mathrm{harmonic}} \simeq \SI{1.286}{\giga\hertz}$, ringing up the tank to the degree where SPAM errors suppress readout entirely.

\section{Conclusions and Outlook}

In conclusion, we have demonstrated all-electrical coherent spin control in the presence of a high-Q superconducting dispersive gate sensor at $\SI{1.5}{\kelvin}$ with a Si FinFET hole serving as a proxy high-density qubit unit cell. We have established that resonator ringing is not detrimental to the qubit coherence but can lead to an increased rate of SPAM errors if the ringing amplitude exceeds the quantum dot's excited state splitting. Note that state leakage out of PSB equally affects all spin readout strategies which rely on spin blockade \cite{Crippa2019,Oakes2023,Pakkiam2018,Niegemann2022,West2019}. Given the strong susceptibility of hole spins to electrical noise \cite{Fang2023}, a robustness to resonator ringing is rather surprising at first glance. Ramsey experiments, however, are particularly sensitive to quasi static noise, while the ringing occurs at several hundred $\SI{}{\mega\hertz}$ thus only weakly impacting $T\mathrm{_2^*}$. Tank ringing also modulates the hole $g$-factor but at a frequency which is too high to affect the spin dynamics due to phase-driving effects \cite{Bosco2023}. Additionally, the double dot level detuning and the tunnel coupling are susceptible to tank ringing which thus directly influences the exchange interaction \cite{Geyer2024}. Such modulations can impact exchange-based two-qubit gates which warrants further investigations.

Typical spin control pulse sequences can ring up the tank circuit's fundamental mode as well as higher harmonics. Harmonics are expected for resonator modes that form between the inductor and the circuit board ground which can likely be addressed by changing the inductor placement. Baseband pulse durations should be chosen such that their harmonics avoid the tank resonance and as long as possible to minimize the amplitude of the harmonic order which lies closest to the tank resonance. We show that the spin coherence and readout efficiency are not affected if these conditions are satisfied. Further solutions may be found by investigating smooth pulse shapes, drive line filtering or varying the phase of subsequent drive pulse repetitions in order to cause destructive interference at the resonator. 

Implementing an on-demand toggle for the tank $Q$ using voltage-tunable capacitances in parallel with the sensor gate \cite{Eggli2023} would be akin to swapping the wire wound surface mount inductor for the superconducting inductor on the experimental time scale. This would require an tunability range from $\simeq\SI{1}{\pico\farad}$ to $\simeq\SI{0.1}{\pico\farad}$. Alternatively, optical light pulses directed at the nanowire inductor could temporarily quench superconductivity, similar to the operation principle of superconducting nanowire single photon detectors \cite{Gol’tsman2001,Erbe2023,EsmaeilZadeh2021} or a transistor in series with the tank could be pulsed to temporarily decouple the tank and the sensor gate \cite{Schaal2018}.

\begin{acknowledgements}
We thank T. Berger, S. Svab, M.J. Carballido and S. Geyer for fruitful discussions and experimental support. We additionally appreciated useful discussions with A. Hamilton, P. Harvey-Collard and M. Mergenthaler. Furthermore, we acknowledge S. Martin and M. Steinacher for technical support. This work was partially supported by the NCCR SPIN, the Swiss Nanoscience Institute (SNI),  Swiss NSF (grant no.\ 179024) and European Union’s Horizon 2020 research and innovation programme under the Marie Skłodowska-Curie grant agreement number 847471.
\end{acknowledgements}

\section*{Data availability}
The data supporting the plots of this work are available at the Zenodo repository at \href{https://doi.org/10.5281/zenodo.11504576}{https://doi.org/10.5281/zenodo.11504576}.

\section*{Author Contributions}
R.S.E, T.P., D.M.Z. and A.V.K. conceived the project and planned the experiments. R.S.E. conducted the experiments, analysed the data and wrote the manuscript with inputs from all authors. The superconducting inductors were designed and fabricated by E.G.K., A.O. and G.S.. D.M.Z., R.J.W. and A.V.K. supervised the project.

\section*{Competing Interests}

The authors declare no competing interests.

\appendix

\section{Drive spectra}
\label{Supp:Spectra}

The resonator spectra depicted in Fig. \ref{fig:1Setup_Spectral_Overlap} were simulated using Qucs, assuming a RLCG transmission line model to account for the higher harmonics with inductor parameters corresponding to the setup of Fig. \ref{fig:4Haronics_Tank}.

\begin{figure}[hbtp]
    %\centering
    \includegraphics[width=0.47\textwidth]{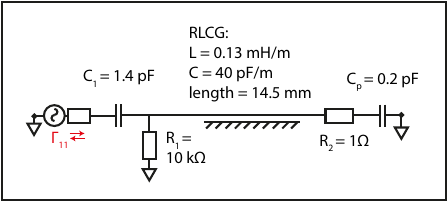}
    \caption{\textbf{Model of high-Q inductor tank circuit} The circuit model features the inductor represented as a RLCG transmission line with negligible resistance, yielding the spectrum presented in Fig. \ref{fig:1Setup_Spectral_Overlap}.
    }
    \label{suppfig:simulation}
\end{figure}

We model the baseband spectrum of Fig. \ref{fig:1Setup_Spectral_Overlap}. (d) by the following series of lorentzian peaks to account for spectral broadening by choosing $\gamma =$ \SI{1}{\mega\hertz} and $A =$ \SI{1}{\mega\hertz} for normalisation to the $n = 1$ peak amplitude:

\begin{equation}
\lvert FFT_{baseband}\rvert (f) =  \sum_{n=1}^{\infty} \frac{A}{2n-1} \frac{\gamma}{\left(f-\frac{2n-1}{2t_{CB}}\right)^2 + \left(\frac{\gamma}{2}\right)^2} 
\end{equation}

The drive burst spectrum in Fig. \ref{fig:1Setup_Spectral_Overlap}. (e) is represented by:

\begin{equation}
    \lvert FFT_{drive}\rvert (f) =  \lvert\frac{\sin(\pi t_{burst} \cdot (f-f_{drive}))}{\pi t_{burst} \cdot (f-f_{drive})}\rvert 
    \label{eq:sinc}
\end{equation}

\section{Data analysis}
\label{Supp:Analysis}

The experimental data for $I_{EDSR}$ and $I_{0,Ramsey}$ as a function of $t_{CB}$ were fitted to functions of the form of equation \ref{eq:I_EDSR}, yielding $\eta_{EDSR}$ and $\eta_{Ramsey}$, respectively as shown in Figs. \ref{fig:3Baseband_Ringup_T2} (c,d,f). For the ringup pattern in Figs. \ref{fig:4Haronics_Tank} (d), we used equation \ref{eq:sinc}.

\subsection{Ramsey fits}
The Ramsey trace in \ref{fig:2LowQ_Tank_Coherence} (c) was fitted according to:

\begin{equation}
\begin{split}
I_{Ramsey} (t_{wait}) = \\
I_{0,Ramsey} \cos(2\pi f_{\phi}t_{wait}+\Phi_0) \exp\left(-\left(\frac{t_{wait}}{T_2^*}\right)^2\right)+I_{offset}
\label{eq:ramsey}
\end{split}
\end{equation}

Here $f_{\phi}$ is the frequency by which the phase of the second $\pi_{\phi}/2$ pulse was varied, with respect to the first $\pi/2$ pulse of the Ramsey sequence, as a function of the Ramsey waiting time $t_{wait}$, $\Phi_0$ is an offset phase factor, the Ramsey current contrast is $I_{0,Ramsey}$, and a finite offset current is captured by $I_{offset}$.
The same model was used in the fits that yielded Fig. \ref{fig:3Baseband_Ringup_T2} (f) and (g). Each trace was recorded 20 times, and the displayed data show the mean of the individual fit results with the standard error.

\subsection{EDSR linewidth fits}

We fit EDSR resonances in the coherence-limited regime to a Gaussian:

\begin{equation}
    I_{EDSR}\left(\Delta f_{drive} \right) = I_{0,EDSR} \exp\left( -\left(\frac{{\Delta f_{drive}}^2}{2\sigma^2}\right)\right)+I_{offset}
\end{equation}
where the EDSR contrast is $I_{0,EDSR}$, the qubit detuning $\Delta f_{drive} = f_{drive} - f_{Larmor}$ is the detuning of the drive frequency relative to the qubits Larmor frequency $f_{Larmor}$, $I_{offset}$ is the background offset current and $\sigma$ is the standard deviation of the Gaussian. We extract $T_2^* = \sqrt{\ln{(2)}}/\left(\pi\sigma\right)$ \cite{Kawakami2014} as a function of $P_{tank}$ for Fig. \ref{fig:2LowQ_Tank_Coherence} (g).

\section{Detailed circuit and experimental setup}
\label{Supp:Circuit}

\begin{figure}[hbtp]
    %\centering
    \includegraphics[width=0.47\textwidth]{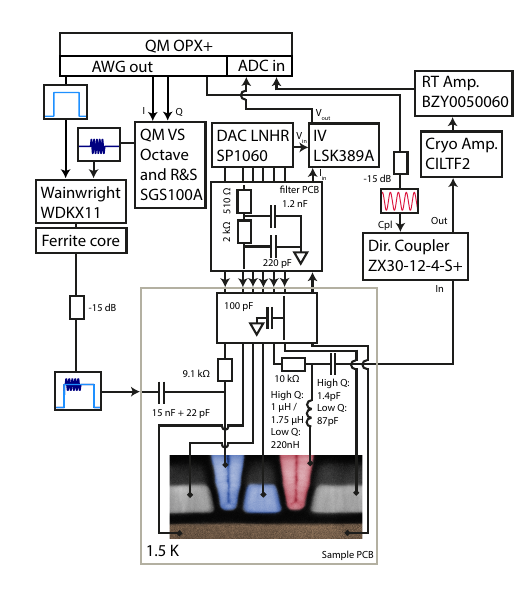}
    \caption{\textbf{Full experimental setup} 
    }
    \label{suppfig2:setup}
\end{figure}

The experimental setup comprised a variable temperature insert with base temperature \SI{1.5}{\kelvin} which hosted a sample- and a filter PCB for signal filtering and combination of DC and RF voltages. The schematic circuit with all discrete elements is shown in \ref{suppfig2:setup}. DC voltages were supplied by a Basel Precision Instruments digital-to-analog converter SP1060 and low-pass filtered on both PCBs. 

The DC current through the FinFET was amplified by a Basel Precision Instruments current-to-voltage converter LSK389A and recorded by a Quantum Machines OPX+ unit. The OPX+ supplied AC signals for qubit control and tank readout. The qubit drive bursts at GHz frequencies were generated by IQ modulation on vector sources (Quantum Machines Octave and, for signals below 2 GHZ, Rohde \& Schwarz SGS 100A). The baseband and control pulses were combined by a Wainwright WDKX11 diplexer and delivered to the sample PCB via attenuated coaxial cables.

The tank readout signal was delivered through the coupling port of a directional coupler (Mini Circuits ZX30-12-4-S+) to the tank resonator on the sample PCB. The reflected signal was amplified at \SI{4}{\kelvin} (Cosmic Microwaves CILTF2) and at room temperature (B\&Z Technologies BZY0050060) and recorded by the second input port of the OPX+.

\section{Qubit measurement scheme}

All qubit measurements followed a scheme where the source-drain current through the FinFET was recorded by the OPX+ input. The main signal was recorded while applying the qubit drive pulse for \SI{5}{\milli\second}, then a reference pulse was applied for \SI{5}{\milli\second} and the reference signal was recorded. This scheme was repeated and the signals were integrated for approximately \SI{1}{\second} for each data point of the qubit measurements (e.g. Rabi, Ramsey and EDSR-traces). Subtracting the reference from the main signal removed a DC offset current and slow drifts of the background. Multiple repetitions of identical scans were averaged to enhance the signal-to-noise ratio e.g.\ in Fig. \ref{fig:3Baseband_Ringup_T2} (c). We now briefly discuss the types of reference signals used for the different experiments.

\subsection{Pulsed measurements}
\label{Supp:Measurement_scheme}

Most pulsed qubit experiments presented in this work are Rabi-like experiments (Rabi chevron in Fig. \ref{fig:2LowQ_Tank_Coherence} (b) and EDSR resonance scans in Figs. \ref{fig:2LowQ_Tank_Coherence} (e), \ref{fig:3Baseband_Ringup_T2} (b,c,d) and \ref{fig:4Haronics_Tank} (b,c,d)). Here, the reference pulse purely consisted of the baseband pulse, removing the qubit drive burst completely. 

The Ramsey experiments were comprised of two $\pi/2$ pulses, separated by the waiting time $t_{wait}$, and shifting the second pulse by a $\phi = 2\pi f_{\phi}t_{wait}$ relative to the phase of the first pulse. This resulted in the sinusoidally modulated current amplitude as a function of $t_{wait}$ as seen in Fig. \ref{fig:2LowQ_Tank_Coherence} (c), improving the fit quality using equation \ref{eq:ramsey}. In order to enhance the readout contrast, the reference sequence for Ramsey experiments was using the same pulse sequence, but shifting the phase of the second pulse by $\phi + \pi$.

\subsection{Continuous drive}
\label{Supp:CW}

The continuous wave experiment in Fig. \ref{fig:4Haronics_Tank} (c) was implemented by only applying the qubit drive for \SI{5}{\milli\second} and recording the leakage current followed by \SI{5}{\milli\second} without high frequency pulsing as the reference.

\section{Additional qubit data}
\subsection{Rabi comparison}
\label{Supp:Rabi_comparison}

\begin{figure}[hbtp!]
    %\centering
    \includegraphics[width=0.47\textwidth]{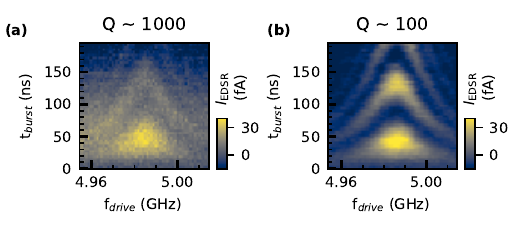}
    \caption{\textbf{Rabi chevrons of device A} \textbf{(a)} recorded with a high-Q resonator connected to P2 and \textbf{(b)} with a low-Q resonator. 
    }
    \label{suppfig3:rabis}
\end{figure}

The quality of the qubit readout is strongly affected by the $Q$ factor of the tank circuit. Fig. \ref{suppfig3:rabis} shows two Rabi chevrons recorded with identical settings using the FinFET device A but in the presence of (a) a high-Q and (b) a low-Q resonator. We furthermore chose $t_{CB}$ such that baseband induced ringup is negligible. While both measurements clearly show a chevron pattern with almost identical Rabi and Larmor frequencies, the contrast in (a) is reduced, overall noise is higher and we observe a shift of the background in (a) as a function of $t_{burst}$ which is independent of $f_{drive}$. This general behavior is reproducible for different readout positions in gate voltage space.

\subsection{Baseband amplitude}
\label{Supp:Baseband_amplitude}

\begin{figure}[hbtp]
    %\centering
    \includegraphics[width=0.47\textwidth]{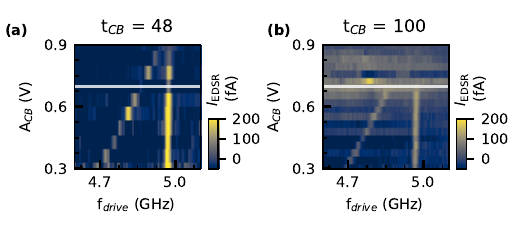}
    \caption{\textbf{Baseband pulse amplitude and ringup} EDSR resonances of the two qubits of device A as a function of the baseband pulse amplitude $A_{CB}$ with \textbf{(a)} $t_{CB} =\SI{48}{\nano\second}$, for which negligible ringup is expected because the harmonics do not match $f_{0}$, and \textbf{(b)} with $t_{CB} =\SI{100}{\nano\second}$, for which ringup is substantial. The resonances vanish in (b) above $A_{CB} =$\SI{0.7}{\volt} (white lines) but prevail in (a). At this point, the ringup amplitude exceeds the excited state level splitting for $t_{CB} =\SI{100}{\nano\second}$ which results in the abrupt loss of readout. 
    }
    \label{suppfig4:Campl}
\end{figure}

From the dependence of the qubit Larmor frequencies on the baseband pulse amplitude $A_{CB}$, we can identify the proximity of the two spins to the driving gate P1 in device A. Fig. \ref{suppfig4:Campl} (a) shows the EDSR resonances of both qubits at fixed magnetic field as a function of $A_{CB}$. The qubit whose resonance shifts strongly with $A_{CB}$ (left resonance) is most likely located under P1 whereas the other qubit is under P2. 

Comparing the spectra for two different $t_{CB}$, we find that the resonances vanish in an increasing background for $t_{CB} =\SI{100}{\nano\second}$ (b), but not for $t_{CB} =\SI{48}{\nano\second}$ (a). In (b), both qubit resonances vanish above the same magnitude of $A_{CB} =$\SI{0.7}{\volt}, consistent with baseband pulse induced ringup. Note that the exact value of $A_{CB}$ at which the contrast vanishes is expected to be different for each $t_{CB}$, depending on the degree to which the baseband harmonics overlap with the tank resonance. The two scans depicted in Fig. \ref{suppfig4:Campl} are illustrative examples, supporting the observation of baseband pulse-induced tank ringing.

\section{Detuning-dependence of high-Q ringup}
\label{supp:detuning_ringup}

In the following, we provide a qualitative explanation of the ringup strength which varies for different $t_{\mathrm{CB}}$ as shown in Fig \ref{fig:3Baseband_Ringup_T2} (d). Overall, the reduction in $I_{\mathrm{EDSR}}$ is stronger if the experimentally accessible value of $t_{\mathrm{CB}}$ matches better one of the odd harmonics of order $n$. 

\begin{figure}[hbtp]
    %\centering
    \includegraphics[width=0.47\textwidth]{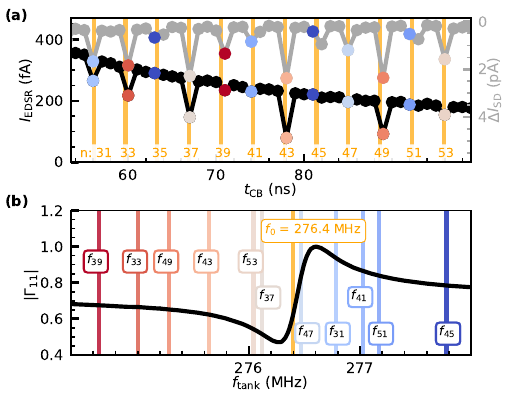}
    \caption{\textbf{Asymmetric response of the high-Q tank to baseband ringup} \textbf{(a)} The EDSR resonance (black) and $\Delta I_{\mathrm{SD}}$ as a function of $t_{\mathrm{CB}}$ is plotted similar to Fig. \ref{fig:3Baseband_Ringup_T2} (d) but over a larger range of $t_{\mathrm{CB}}$. The colored datapoints are the closest experimentally accessible $t_{\mathrm{CB}}$ to the respective harmonic (orange) of order $n$. \textbf{(b)} shows the tank resonance (black) and calculated frequencies $f_{\mathrm{n}} = n f_{\mathrm{bb}}$ with the colors corresponding to (a). Ringup-induced dips in $I_{\mathrm{EDSR}}$ are deeper for red detuned cases ($f_{\mathrm{n}}<f_{\mathrm{0}}$) than for blue detuned frequencies ($f_{\mathrm{n}}>f_{\mathrm{0}}$). The asymmetry of the tank resonance explains this finding: The resonator is more easily excited by slight red detuned frequencies whereas its reflectance is enhanced for blue detuning.
    }
    \label{suppfig5:Detuning}
\end{figure}

Additionally, the depth of the ringup-induced dips depends on the detuning of the applied baseband pulse harmonic with respect to $f_{\mathrm{0}}$ as shown in Fig. \ref{suppfig5:Detuning}. Because the tank resonance is asymmetric, red-detuned baseband harmonic excitations ($f_{\mathrm{n}}<f_{\mathrm{0}}$) cause much stronger ringup than blue-detuned frequencies ($f_{\mathrm{n}}>f_{\mathrm{0}}$). Such asymmetric tank resonances have been observed in previous studies with superconducting high-Q inductors \cite{Kelly2023, Ibberson2021} and are attributed to imperfect impedance matching of the circuit.

\end{document}